\documentclass[a4paper,11pt]{amsart}
\begin{document}

\title{{\bf  ``QUANTUM GRAVITY'': AN OXYMORON}}
\author{Angelo Loinger}
\date{}
\address{Dipartimento di Fisica, Universit\`a di Milano, Via
Celoria, 16 - 20133 Milano (Italy)}
\email{angelo.loinger@mi.infn.it}
\thanks{To be published on \emph{Spacetime \& Substance}}

\begin{abstract}
I prove that ``quantum'' and ``Einsteinian gravity'' are
incompatible concepts. Accordingly, the graviton is a mere object
of scien\-ce fiction.
\end{abstract}

\maketitle

\vskip1.20cm
\noindent{\bf 1.}-- {\bf Introduction}\par \vskip0.10cm The
innumerable and learned efforts during seventy years to create a
quantum formulation
 of general relativity have only beaten the air -- \emph{et pour cause}, as we shall see.
 On the other hand, it is evident to any unprejudiced scientist that definite reasons must
 be at the root of this failure.\par
 First of all, whereas ``particles and fields exist \emph{within}
 space-time, gra\-vi\-ty \emph{is}, in essence, space-time'' \cite{1}. This
 implies, in particular, that the physical meaning of the so-called
 critical (or ``Planckian'') quantities
 $M_{0}\equiv(\hbar c/G)^{1/2}\approx10^{-5}\ \textrm{g}$,
 $L_{0}\equiv(\hbar/M_{0}c)\approx10^{-33}\ \textrm{cm}$ and $T_{0}\equiv
 L_{0}/c$ is rather uncertain (``unsicher''), as it was emphasized
 by Rosenfeld many years ago \cite{2}. Rosenfeld was specially
 qualified to formulate a judgment of that kind because the above
 constants came forth through an extension to the quantized
 \emph{\textbf{linear}} approximation of general relativity (whose
 substrate is Minkowski spacetime -- and this is an essential
 point) of a deep method, created by Bohr and Rosenfeld for the
 quantum electromagnetic field \cite{3}.\par
The current belief that below time $T_{0}$, length $L_{0}$, and
mass $M_{0}$ the Einsteinian theory of gravitation loses its
validity is fully unfounded. Indeed, its justification by means of
a bold application of more or less sophisticated quantum
techniques does not possess any sound basis. General relativity
has nothing to do with the classical field theories in Minkowski
spacetime, or in ``rigid'' Riemann-Einstein spacetimes.\par
Further, ``there is no experiment that tells us that the
quantization of gravity is necessary'' \cite{1}.\par Finally, the
fictive nature of the so-called gravitational waves \cite{4} is
sufficient to render meaningless \emph{any} quantization program
of general relativity.\par
 (The physical inconclusiveness of the
theoretical approaches that make use of supplementary dimensions
of spacetime curled up with a radius comparable to the
``Planckian'' length $L_{0}$, does not need to be emphasized. The
opinion according to which the superstring theory provides a
possibility for a consistent quantum theory of gravity is
destitute of a rational foundation).\par
 For a bibliography on quantum gravity see e.g. the \emph{References} of
the papers \cite{5} and \cite{6}.

\vskip0.80cm
\noindent {\bf 2.}-- {\bf Two observational results} \par
\vskip0.10cm Two recent papers of \emph{observational} nature (see
\cite{5} and \cite{6}) raise serious doubts on the existence of
the quantum fluctuations of the metric tensor of general
relativity at the ``Planckian'' scales, i.e. at the scales of the
constants $L_{0}$, $M_{0}$, $T_{0}$ (see sect.{\bf 1}).\par
 Lieu
and Hillman \cite{5}  remark that if the above fluctuations really
existed, the instant $t$ of an event could not be determined more
accurately than a standard deviation
$\sigma_{t}/t=a_{0}(T_{0}/t)^{\alpha}$, where $a_{0}$ and $\alpha$
are positive constants $\sim1$. (Analogously, the distances should
be subject to an ultimate uncertainty $c\sigma_{\tau}$.) As a
consequence of a cumulative effect of this ``Planck-scale
phenomenology'', we should have a complete loss of phase of the
e.m. radiation emitted at large distances from the observer. The
conclusion of the abstract of paper \cite{5} runs as follows:
``Since, at optical frequencies, the phase coherence of light from
a distant point source is a ne\-ces\-sa\-ry condition for the
presence of diffraction patterns when the source is viewed through
a telescope, such observations offer by far the most sensitive and
uncontroversial test. We show that the HST [Hubble Space
Telescope] detection of Airy rings from the active galaxy
PKS1413+135, located at the distance of $1.2$ Gpc secures the
exclusion of all first order ($\alpha=1$) quantum gravity
fluctuations with an amplitude $a_{0}>0.003$. [\ldots]''\par
Ragazzoni, Turatto and Gaessler \cite{6} write: ``[\ldots] We
elaborate on such an approach [i.e., the approach of \cite{6},
which was subject to some criticism] and demonstrate that such an
effect would lead to an ap\-pa\-rent blurring of distant point
sources. Evidence of the diffraction pattern from the HST
observations of SN1994D and the unresolved appearance of a Hubble
Deep Field galaxy at $z=5.34$ lead us to put stringent limits on
the effects of Planck-scale phenomenology.''\par I shall now prove
rigorously that, from a sound theoretical standpoint, there are
\emph{no} quantum fluctuations of the fundamental tensor of
general relativity.

\vskip0.80cm
\noindent {\bf 3.}-- {\bf The oxymoron}
\par \vskip0.10cm As is was explicitly pointed out by Pauli \cite{7}, in quantum
mechanics the time $t$ is a ``\emph{gew\"ohnliche Zahl}
($_{``}$\emph{c-Zahl}'')'', i.e. it coincides with the time of
classical physics. Thus, time $t$ is \emph{not} a dynamical
variable represented by an operator of the Hilbert space of the
physical states. Analogously, also the co-ordinates of the points
of three-dimensional physical space are parameters, and not
dynamical variables; only the co-ordinates $q_{r}$,
$(r=1,2,\ldots, n)$ of the $n$ degrees of freedom of a holonomic
system are dynamical variables represented by Hilbert
operators.\par
 In the customary (Lorentzian) quantum field theory,
a given field is described by a set of $m$, say, operators
$\varphi_{s}$, $(s=1,2,\ldots,m)$, that are functions of the
spatial points and of the instants of time.\par In \emph{general}
relativity the fundamental spacetime interval $\textrm{d}s$ is
given by
\begin{equation}\label{1}
\textrm{d}s^{2}=g_{jk}(x^{0},x^{1},x^{2},x^{3})\textrm{d}x^{j}\textrm{d}x^{k},
(j,k=0,1,2,3) \:\:\:,
\end{equation}
where the coefficients $g_{jk}$ of the metric do \emph{not}
represent a classical field in the conventional meaning, but
characterize directly the spatiotemporal structure -- in other
terms, \emph{they ``are'' the spacetime itself}. (The co-ordinates
$x^{0}$, $x^{1}$, $x^{2}$, $x^{3}$ are mere \emph{labels} of the
spacetime points, fully devoid of any metrical meaning).\par
 If we
write the Minkowskian $\textrm{d}s^{2}$ of a Lorentzian quantum
theory making use of a system of general co-ordinates $x^{0}$,
$x^{1}$, $x^{2}$, $x^{3}$, we obtain obviously an expression of
the following kind:
\begin{equation}\label{2}
\textrm{d}s^{2}=h_{jk}(x^{0},x^{1},x^{2},x^{3})\textrm{d}x^{j}\textrm{d}x^{k},
(j,k=0,1,2,3)\:\:\:,
\end{equation}
and we see that, according to the basic axiom previously
emphasized \cite{7}, the functions $h_{jk}(x)=h_{kj}(x)$ are
non-operators, i.e. they are (necessarily!) customary functions
(``c-numbers'') of the co-ordinates $x^{0}$, $x^{1}$, $x^{2}$,
$x^{3}$.\par We realize now that the project of a theory such that
the $g_{ik}$'s of the \emph{exact} (non-approximate) formulation
of \emph{general} relativity are promoted to the role of operators
of a function space implies a blatant contradiction with the above
axiom of quantum theory.\par ``Quantum'' and ``[Einsteinian]
gravity'' are incompatible concepts, and thus the expression
``quantum gravity'' is actually an oxymoron.

\vskip0.80cm
\noindent {\bf 4.}- {\bf Recapitulation}
\par \vskip0.10cm
The classic spacetimes of quantum theories are the following:
\emph{i}) the Euclidean-Newtonian substrate of Galilean group of
transformations; \emph{ii}) the Minkowskian substrate of
Lorentzian group of transformations; \emph{iii}) any given,
``rigid'' Riemann-Einstein spacetime.\par We have correspondingly:
\emph{i}) the nonrelativistic quantum mechanics of the systems
with a finite number of degrees of freedom; \emph{ii}) the
Lorentzian quantum theories -- and the quantized \emph{linear}
approximation of GR (Pauli, Rosenfeld); \emph{iii}) Dirac's
equation for a particle in a fixed Riemann-Einstein spacetime.\par
The known quantum formalisms can have a definite physical sense
\emph{only} under the condition that the above spacetimes are
described by the cus\-to\-ma\-ry non-operator entities.
Consequently, the meaning of \emph{any} quantization program of GR
is doomed to a whimsical arbitrariness, because it implies
necessarily some \emph{operator} characterization of spacetime
itself.

\vskip0.80cm
\noindent APPENDIX\\ \emph{\textbf{A puffing operation}}
\par \vskip0.10cm As it has been recalled in sect.\textbf{1}, the constants $L_{0}$,
$M_{0}$, $T_{0}$ pertain, ri\-go\-rous\-ly speaking, \emph{only}
to the quantum \emph{linearized} version of GR. In the current
astrophysical literature they are denominated ``Planck
constants''. Why? The reason is simple. In 1899 Planck \cite{8}
remarked that with suitable combinations of the \emph{fundamental}
constants $G$, $c$, $h$, it is possible to obtain the following
four ``natural'' \emph{units of measure}:\\
unit of length: $\sqrt{\frac{Gh}{c^{3}}}$ ,\\
unit of mass: $\sqrt{\frac{ch}{G}}$ ,\\
unit of time: $\sqrt{\frac{Gh}{c^{5}}}$ ,\\
unit of temperature: $\frac{1}{k}\sqrt{\frac{c^{3}h}{G}}$ ,\\
where $k$ is Boltzmann's constant. (Actually, in the paper of 1899
Planck wrote $b$ in lieu of $h$, and $a$ in lieu of $h/k$.)\par
Clearly, ``measure units'' and ``physical constants'' are
\emph{distinct} concepts. I suppose that the astrophysical
community is perfectly aware of this trivial difference.\par To
qualify with Planck's name the constants $L_{0}$, $M_{0}$, $T_{0}$
has been a tricking operation with the aim to dignify with a great
name three constants having a very dubious meaning.

 \small \vskip0.5cm\par\hfill{\emph{``}\emph{--Warum willst du
dich von uns allen}
  \par\hfill \emph{Und unsrer Meinung entfernen?--}
  \par\hfill \emph{Ich schreibe nicht euch zu gefallen,}
  \par\hfill \emph{Ihr sollt was lernen!''}
  \vskip0.10cm\par\hfill \emph{J.W.v. Goethe}}

\end{document}